# Designing Visual Learning Analytics for Supporting Equity in STEM Classrooms

Ali Raza*, William R. Penuel, Tamara Sumner

Department of Computer Science, Institute of Cognitive Science, University of Colorado Boulder

**ABSTRACT**

Supporting equitable instruction is an important issue for teachers attending diverse STEM classrooms. Visual learning analytics along with effective student survey measures can support providing on time feedback to teachers in making instruction more culturally relevant to all students. We adopted a user-centered approach, where we engaged seven middle school science teachers in iterative testing of thirty data visualizations disaggregated over markers such as gender and race for implementation of selected displays in a visual learning analytics tool- Student Electronic Exit Ticket (SEET). This process helped us gather insights into teachers' sense-making in identifying patterns of student data related to gender and race, selecting and improving the design of the feedback displays for the SEET [10].

## 1  INTRODUCTION

Understanding equity in classrooms using visual analytics is in the nascent phase of supporting diversity across gender and race [2]. Whereas improving classroom learning using analytics is not a new dimension and as Siemens [1] explained analytics and big data are going to play a guiding role in reforming and shaping higher education and also how they can assist learning and teaching in education. Supporting equitable learning is a critical issue in STEM classrooms especially when the classes have diverse students.

We are leveraging visual analytics as a tool for helping teachers reflect on the equity issues in STEM classrooms. For designing visual feedback displays that are aligned with the visual sense-making ability of the teacher, we adopted a user-centered design approach. As seen in prior work by Zack & Franconeri [3], instead of following defined guidelines in designing visualizations for changing datasets in different contexts, it is critical to experiment, gather feedback, and iterate on the visualizations. We conducted two study iterations with seven science teachers. For the first iteration, we prepared ten possible visualizations from a real classroom dataset collected in a science classroom on measures related to coherence, relevance, and contribution which are found to reliably capture student experience of a particular lesson [4]. We gathered feedback from four science teachers using think aloud and followed by a cognitive interview. Based on the feedback from this iteration, we designed twenty more visualizations from the same dataset and tested the highest-rated visualizations from iteration one and newly designed with three more science teachers.

* Email: a.raza@colorado.edu

For this iteration, we adopted the same methodological approach from the iteration one. Analysis on the completion of these two design cycles and along with the research team observations helped in finalizing: bar chart, heat map and line chart for displaying disaggregated and over time data by gender and race (Figure 1, 2, 3).

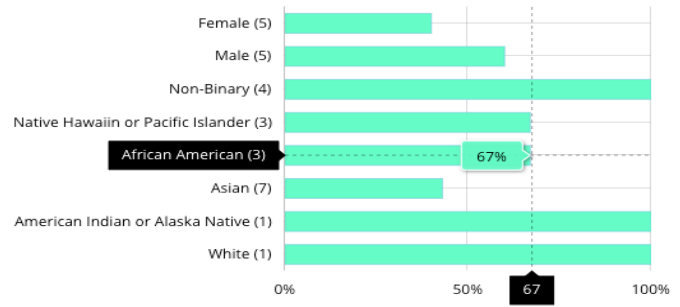

Figure 1: Bar chart disaggregated by gender and race for a single question

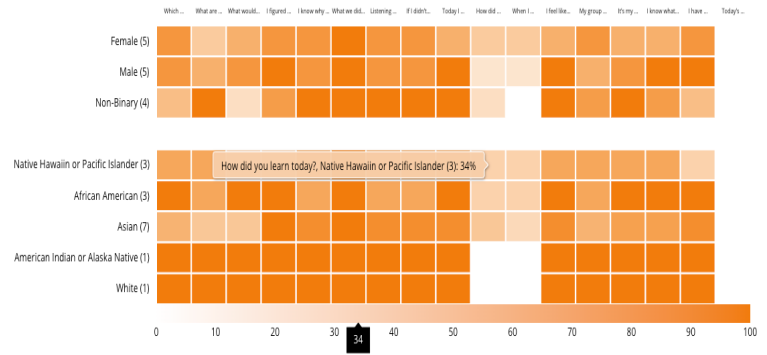

Figure 2: Heat map disaggregated by gender and race. Mouse over displays the question, student gender or race, and percentage of students who said 'yes' to the question

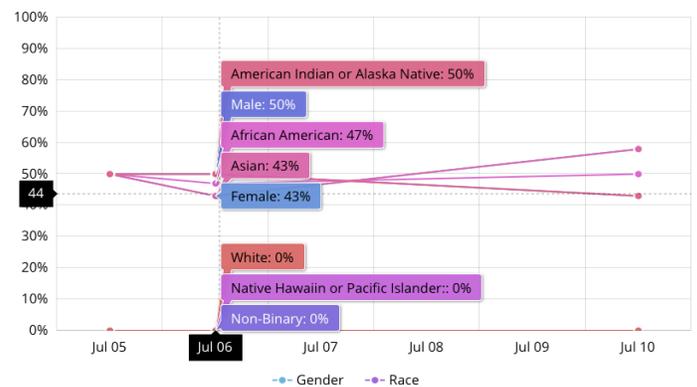

Figure 3: Line chart displaying information by gender and race with interactive legends. If clicked on legend below, it displays information only for gender or race

## 2 THEORY

Researchers from the Learning analytics (LA) community emphasized the importance of human-centered methods for designing analytics dashboards. Similarly, we adopted a human-centered approach by involving teachers in the design process with different data visualization displays. Zacks & Tversky [5] in their classic experiment narrated how changing different visualization displays change the questions asked by the participants involved in the experiment. In our process of capturing, data cleaning, and visualizing, all questions belonged to three broad categories. Coherence asks students whether they understand how current classroom activities contribute to the larger investigations in which they are engaged. Relevance questions ask students to consider the degree to which lessons matter to students themselves, to the class, and the larger community. For contribution, the SEET survey asks students whether they shared their ideas in a group discussion, heard ideas shared by others, and whether others' ideas impacted their thinking [4]. We created displays based on these three broad categories.

## 3 RESULTS

Results from the two design sessions with teachers highlighted significant strengths and weaknesses in different visualizations. In the process, we also received design suggestions from teachers that supported in selecting visualizations for implementing in the SEET tool. From both sessions, all teachers selected bar charts, heat map, and line charts for displaying student experience data based on gender and race. Figures 1, 2, and 3 illustrate all three visualizations, the bar chart is displaying disaggregated data for gender and race for one question related to a particular measure. We implemented small multiples of bar charts, adjacent to each other as suggested in prior work by Heer, Bostock & Ogievetsky [6]. This allowed us to create an effective visualization experience for users. The heat map in Figure 2 is displays data disaggregated by gender and race, the first three rows at the top display different genders, and then there is intentional space to start with different race categories. These categories are based on the U.S. census. Each column at the top belongs to lesson or measures (coherence, relevance, and contribution) related questions. When a user mouse over a matrix of the heat map, they can view the question, gender or race, percentage correct option for the particular box. We are leveraging the power of the human visual system with the use of color encoding to help recognize different patterns in the data. Although the color is less precise as compared to position based visual representations such as scatterplots, bar graphs, and line graphs. But they provide a significant 'big picture' information to the users [7]. The line chart in Figure 3 is one type of implementation of its three variants: i) line chart visualized by measures only questions over time ii) line chart visualized by gender and race data over time for a single classroom period (see Figure 3) iii) line chart visualized by each question over time for a single classroom period. All these visualizations and variants are implemented in the SEET system for supporting equity in STEM classrooms.

## 4 DISCUSSION

Using visual learning analytics in supporting learning is not a new endeavor. It has been found that learning analytics is useful for driving instruction and is supportive of individual students' needs. Although using analytics to support equity is in a growing stage, as our classrooms become more diverse culturally, it is critical to understand the individual experience of the students and make a shift towards cultural relevant pedagogy. In our work, we are conjecturing to use visual learning analytics using our SEET platform to support equity in K-12 STEM classrooms. We are using surveys as practical measures in three broad categories; coherence, relevance, and contribution, as a key to understanding student experience in the form of formative assessments in the classroom. They are "practical" in that they can be collected, analyzed, and used within the daily work lives of practitioners. They are also "practical" in that they measure practice [8]. A high-level SEET architecture user flow looks as: A teacher launches the survey measure for that class, and students complete the survey. Then, the student data from the survey is visualized on a teacher facing dashboard. The teacher then engages in a sense-making process to learn about the classroom experience of the students from the dashboard and finally identifies action based on the feedback received from classroom.

How people think visually and make decisions using visualization is being immensely studied. Still, we need to keep up with the pace of ever-changing complex datasets in all domains of study. Wall and colleagues [9] provided a design space for mitigating cognitive bias in visualizations; two ways that can mitigate the bias in visualization are visualization representation and interaction design. In our context of working with the middle school science teachers, we explored their sense-making related to different visual feedback displays and identified ones that helped them effectively notice patterns of data related to inequity in a STEM classroom. There is an emergent need to study teachers' perception and cognitive understanding of visualization and how they make instructional or curricular decisions using a wide variety of data to improve their classrooms.